\documentclass[aps,prl,twocolumn,groupedaddress,reprint]{revtex4-1}
\usepackage{graphicx}
\usepackage{setspace}
\usepackage{amsmath}
\usepackage{amssymb}
\usepackage{color}
\usepackage{array}
\usepackage{subfigure}
\usepackage{hyperref}
\usepackage{float}
\usepackage{lipsum}

\usepackage[all]{xy}
\newcommand{\RN}[1]{%
  \textup{\uppercase\expandafter{\romannumeral#1}}%
}

\DeclareMathOperator{\Tr}{Tr}

\begin{document}
\title{Extending frequency-bin entanglement from photon-photon to atom-photon hybrid systems}
\author{Yuan Sun$^1$}
\email[email: ]{sunyuan17@nudt.edu.cn}
\author{Chang Liu$^2$}
\author{Bao-Quan Ou$^1$}
\author{Ping-Xing Chen$^1$}

\affiliation{$^1$ Interdisciplinary Center for Quantum Information, National University of Defense Technology, Changsha 410073, P.R.China}

\affiliation{$^2$ Institute for Quantum Computing and Department of Physics \& Astronomy, University of Waterloo, Waterloo, Ontario N2L3G1, Canada}

\begin{abstract}
Inspired by the recent developments of atom-photon quantum interface and energy-time entanglement between single-photon pulses, we propose a viewpoint of the single-atom single-photon entanglement based on energy-time considerations, which is analogous to the frequency-bin entanglement between single-photon pulses. We show that such entanglement arises naturally in considering the interaction between frequency-bin entangled single photon pulse pair and a single atom, via straightforward atom-photon phase gate operations. Its anticipated properties and a preliminary example of its potential application in quantum networking are also demonstrated. Moreover, we construct convenient quantum entanglement witness tools to detect such extended frequency-bin entanglement from a reasonably general set of separable states.
\end{abstract}
\maketitle

Manipulating the interaction between a single atom and a single-photon optical pulse is of essential importance in the research frontier of quantum physics. In particular, a lot of efforts have been devoted to studying the single-atom single-photon entanglement, which enables the potential of establishing a large scale quantum network for quantum communication and quantum computing \cite{Kimble07127, RevModPhys.82.1209, PhysRevA.89.022317, RevModPhys.87.1379}. Many exciting progresses have been achieved so far, including the deterministic controlled phase gate between a trapped single atom and a flying single photon pulse via cQED method \cite{PhysRevLett.92.127902, Reiserer13177, Reiserer1349} and the fast quantum gate between ion qubit and frequency encoded photonic qubit \cite{PhysRevLett.97.040505}. These efforts culminated with the advent of the experimental demonstration of quantum networking between matter qubits, especially the cold atoms in optical cavities \cite{Ritter12} and trapped ions \cite{Olmschenk486, PhysRevLett.102.250502}.
\par
Meanwhile, energy-time entanglement between photon pulses \cite{nphys2492, PhysRevLett.82.2594, PhysRevLett.84.5304}, has attracted more and more attention in recent years as the experimental techniques keep on improving \cite{PhysRevLett.113.063602, PhysRevLett.118.030501}. It enhances our understanding of quantum physics fundamentals with providing convincing evidence against the local hidden variable theory \cite{PhysRevA.47.R2472, PhysRevLett.66.1142, PhysRevLett.93.010503, nature07121, PhysRevLett.110.260407}, and has practical applications in quantum information, such as the quantum cryptography and the quantum key distribution \cite{PhysRevLett.84.4737, PhysRevLett.98.060503, PhysRevLett.112.120506, 1367-2630-16-1-013033}, ever since the proposal of Franson interferometer \cite{PhysRevLett.62.2205}. The advantage of energy-time entanglement is its large information capacity, highly non-local properties, and compatibility with nowadays fiber optic technology infrastructure, which have been demonstrated by various interferometric methods \cite{Stucki2005a, PhysRevA.87.053822, PhysRevA.91.053851,  PhysRevLett.65.321, PhysRevLett.81.3563, PhysRevLett.101.180405, PhysRevLett.102.040401}. Two special versions of energy-time entanglement, the time-bin entanglement and frequency-bin entanglement, are of particular practical interest due to their friendliness to experiments; for example, the time-bin entanglement can be stored and retrieved in quantum memory \cite{Gisin09662}, while  the frequency-bin entanglement can utilized to construct the biphoton frequency comb \cite{Xie2015110}. 
\par
Probabilistic entanglement between frequency encoded photonic qubit and ion qubit has already been experimentally demonstrated \cite{PhysRevLett.97.040505} by employing two-photon interference \cite{PhysRevA.85.021803}. Moreover, very recent experimental progress has demonstrated the potential of frequency-encoded photonic qubits in quantum information processing and quantum communication \cite{PhysRevLett.117.223601}. Intuitively, inspired by the success of the energy-time entanglement between single photon pulses and the single-atom single-photon coupling technique, one may wonder whether a parallel concept of energy-time entanglement can be established between a single-photon pulse and a single atom. This question motivates our work, and we hope our work may further assist the efforts along the direction of building quantum correlations via frequency-bin encoded photon pulses \cite{Lukens:17}.
\par
In this letter, we are trying to adopt the frequency-bin entanglement viewpoint to treat the entanglement between a frequency encoded single photon pulse and an atom's internal electronic states \cite{SuppInfo}. We demonstrate how this entangled state can be generated via the interaction between a frequency-bin entangled photon pulse pair and a single atom via a controlled-Z (C-Z) atom-photon quantum phase gate. Then we extend the concept to multi-component case and discuss the entanglement witnessing method. The proposed theory is in principle applicable to a variety of qubit platforms interacting with light in the optical wavelength range, such as neutral atoms, ions, quantum dot and color centers in crystal, even though it is presented here in the setting of neutral atoms.
\begin{figure}[t!]
 \centering
\begin{tabular}{l}
\includegraphics[trim = 0mm 0mm 0mm 0mm, clip, width=8.5cm]{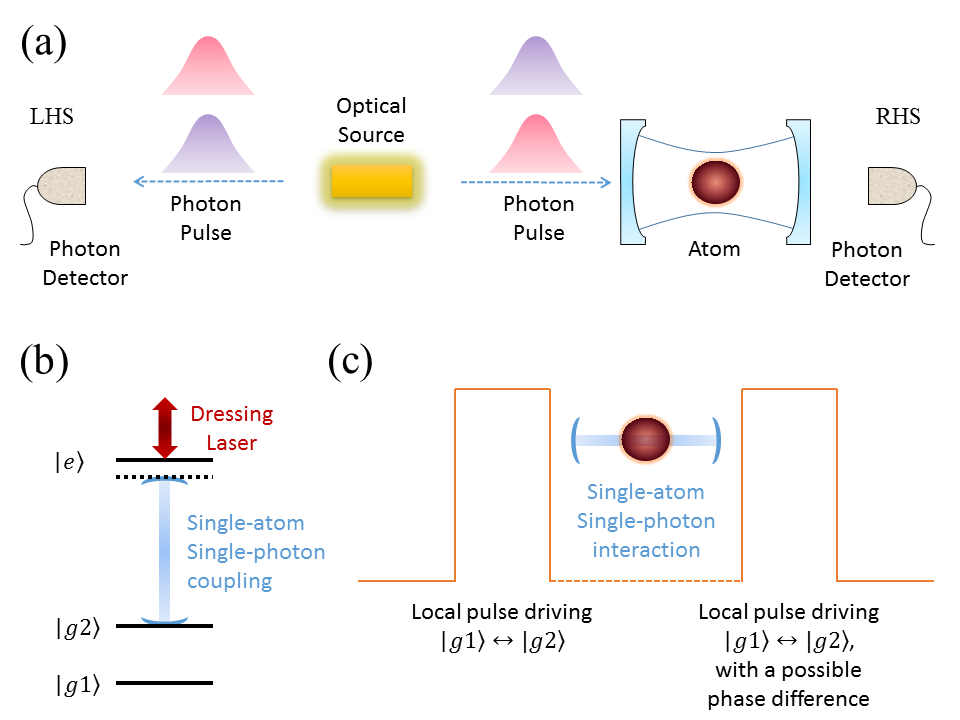}
\end{tabular}
\linespread{1} 
\caption{(a) Schematic of the atom-photon interaction. The optical source emits a pair of energy--time entangled single photon pulses which travel to the right hand side (RHS) and left hand side (LHS) respectively. On RHS, the single photon pulse interacts with the atom via strong atom-photon coupling. (b) Idealized energy level structure of the atom, where $|e\rangle$ is the excited state whose exact energy receives fine-tuning by a dressing laser via ac Stark shift. For example, $|g1\rangle, |g2\rangle$ can be recognized as the states $|F=1, m_F=0\rangle, |F=2, m_F=0\rangle$ of the Na atom ground hyperfine levels.  (c) On RHS, the pulse sequence to transcribe the energy-time entanglement from photon pulse to atom, where the single-atom single-photon interaction effectively serves as a C-Z gate.\label{schematic_1}}
\end{figure}
\par
More specifically, we are going to study frequency-bin type atom-photon entanglement, such as the following hybrid correlated system, consisting of photonic frequency-resolved states and atom's internal electronic states:
\begin{equation}
\label{FBE_2comp_triplet_+1}
|\Psi_2\rangle = \frac{1}{\sqrt{2}}
(|\omega_4\rangle_\text{L} 
|g2\rangle_\text{R} +
 |\omega_3\rangle_\text{L} 
|g1\rangle_\text{R}),
\end{equation}
where the subscript L stands for LHS and the subscript R stands for RHS. It is analogous to the two-component frequency-bin entanglement of narrow-band photon pulses:
\begin{equation}
\label{FBE_2comp_photons}
|\Phi_2\rangle = \frac{1}{\sqrt{2}}
(|\omega_4\rangle_\text{L} 
|\omega_2\rangle_\text{R} +
 |\omega_3\rangle_\text{L} 
|\omega_1\rangle_\text{R}),
\end{equation}
where for simplicity we assume that $\omega_4 + \omega_2 \approx \omega_3 + \omega_1$, therefore the phase difference over time can be precisely and practically traced over the interaction time duration, and that $\omega_1$ is close to the frequency of the transition $|g2\rangle \leftrightarrow |e\rangle$.
\par
As a first step, we show the procedure to generate the state of Eq. \eqref{FBE_2comp_triplet_+1} from the state of Eq. \eqref{FBE_2comp_photons} via suitable interaction with a single atom. The interaction process is sketched in Fig. \ref{schematic_1}(a), the internal level structure of the atom is sketched in Fig. \ref{schematic_1}(b) and the pulse sequence is sketched in Fig. \ref{schematic_1}(c). This is just one out of many possible methods to generate the state of Eq. \eqref{FBE_2comp_triplet_+1}, yet it in principle allows the photon pulse on LHS to be of virtually any color prescribed by the photonic state $|\Phi_2\rangle$.
\par
Suppose the atom is prepared in the state $|g1\rangle$ initially upon the incident photon pulse, where the state of the entire system is
\begin{equation}
\label{1atom_2spp_initial1a}
\frac{1}{\sqrt{2}}(|\omega_4\rangle_\text{L} |\omega_2\rangle_\text{R} 
+ |\omega_3\rangle_\text{L} |\omega_1\rangle_\text{R}) 
\otimes
|g1\rangle_\text{R},
\end{equation}
which is up to a global phase that does not matter. After applying a local $\pi/2$-pulse to the $|g1\rangle \leftrightarrow |g2\rangle$ transition of the atom, the state becomes: 
\begin{equation}
\label{1atom_2spp_initial1b}
\frac{1}{\sqrt{2}}(|\omega_4\rangle_\text{L} |\omega_2\rangle_\text{R} 
+ |\omega_3\rangle_\text{L} |\omega_1\rangle_\text{R}) 
\otimes
\frac{1}{\sqrt{2}} (|g1\rangle_\text{R} - i|g2\rangle_\text{R}),
\end{equation}
where $|g1\rangle, |g2\rangle$ are in the rotating wave frame defined by the frequency of the local pulse $\omega_{g2} - \omega_{g1}$.
Then follows the atom-photon C-Z gate on RHS between the single atom and the incident photon pulse, where the combination of $|\omega_1\rangle_\text{R}$ and $|g2\rangle$ will incur an additional $\pi$ phase shift. Afterwards the RHS photon pulse is to be projected onto the time eigenstate basis -- registered as a click at time $t_\text{R}$ on the photon detector which has a frequency response range larger than $|\omega_1 - \omega_2|$. In other words, the RHS photon pulse is serving as the heralding signal, and the heralded state consisting of the LHS photon and RHS atom is 
\begin{align}
\label{1atom_2spp_heralded1a}
|\Psi_{1h}\rangle = \frac{1}{\sqrt{2}}
\{ e^{-i\omega_2 t_\text{R}} |\omega_4\rangle_\text{L} \cdot
\frac{1}{\sqrt{2}}(|g1\rangle_\text{R} - i|g2\rangle_\text{R}) 
\nonumber\\
- e^{-i\omega_1 t_\text{R}} |\omega_3\rangle_\text{L} \cdot
\frac{1}{\sqrt{2}}(|g1\rangle_\text{R} + i|g2\rangle_\text{R}) \}.
\end{align}
Eventually a second local $\pi/2$-pulse can be applied to the RHS atom, which will change the state of Eq.\eqref{1atom_2spp_heralded1a} into: 
\begin{equation}
\label{1atom_2spp_heralded1b}
\frac{1}{\sqrt{2}}
(-ie^{-i\omega_2 t_\text{R}} |\omega_4\rangle_\text{L}|g2\rangle_\text{R} - 
e^{-i\omega_1 t_\text{R}} |\omega_3\rangle_\text{L}|g1\rangle_\text{R}),
\end{equation}
which is almost the same as Eq.\eqref{FBE_2comp_triplet_+1}, apart from a relative phase.
\par
One essential step leading to the state of Eq.\eqref{1atom_2spp_heralded1b} is the atom-photon C-Z gate on RHS. If we restrict the system to be a single neutral atom trapped inside a high Q optical cavity where the incident photon is to be ultimately reflected from the cavity, then the phase gate design of Ref. \cite{PhysRevLett.92.127902} suffices for this purpose \cite{SuppInfo}, which has been demonstrated successfully in several recent experiments \cite{Reiserer1349, Reiserer13177, Chen768}. We also note several recent developments that will enhance the atom-photon coupling \cite{PhysRevLett.113.133601, PhysRevLett.114.220501, PhysRevA.94.053830}. This type of C-Z gate is a natural consequence of the interaction between the atomic transition $|g2\rangle \leftrightarrow |e\rangle$ and cavity mode described by the following Jaynes-Cummings Hamiltonian: 
\begin{equation}
\label{CZ_JCM1}
\mathbf{H}_\text{int} 
=  \frac{\hbar \Omega_1}{2}( |e\rangle \langle g2| \hat{a}_c 
+ |g2\rangle \langle e| \hat{a}^\dagger_c),
\end{equation}
where the standard input-output relation holds for the incidence field $\hat{a}_\text{in}$, the output field $\hat{a}_\text{out}$ and the cavity field $\hat{a}_c$: $\hat{a}_\text{out} = \hat{a}_\text{in} + \sqrt{\kappa} \hat{a}_c$ with $\Omega_1$ being the single-atom single-photon coupling and $\kappa$ being the cavity linewidth \cite{PhysRevLett.92.127902, SuppInfo}. When the incident photon pulse is of frequency $\omega_2$, it is so far off-resonant such that it is reflected without entering the cavity. When the incident photon pulse is of frequency $\omega_1$, if the atom is at state $|g2\rangle$, the cavity resonance frequency is effectively shifted therefore the incident photon pulse still gets reflected without entering the cavity, while if the atom is at state $|g1\rangle$ the incident photon pulse is resonant with the cavity and actually enters.
\par
Ideally, via repeated applications of the atom-photon C-Z gate, an optical state of Eq. \eqref{FBE_2comp_photons} is capable of entangling two distant atoms of the same species under the condition of $\omega_4=\omega_1, \omega_3=\omega_2$. This can also be regarded as a quantum networking via energy-time entangled photon pulses. More specifically, consider the initial state of:
\begin{equation}
\label{2atom_2spp_initial1a}
|g1\rangle_\text{L}|g1\rangle_\text{R}
\otimes
\frac{1}{\sqrt{2}}(|\omega_1\rangle_\text{L} |\omega_2\rangle_\text{R} 
+ |\omega_2\rangle_\text{L} |\omega_1\rangle_\text{R}),
\end{equation}
where the entire system is made up by two atoms and two photon pulses. The basic operation sequence is conforming to the same principle as Fig. \ref{schematic_1}. \RN{1}: apply local $\frac{\pi}{2}$-pulses on LHS and RHS atoms respectively, to prepare the atoms in the superposition state of $|g1\rangle$ and $|g2\rangle$ in the same manner as Eq. \eqref{1atom_2spp_initial1b}; \RN{2}: apply the atom-photon C-Z gate on both LHS and RHS; \RN{3}: project both single photon pulses on to time basis to obtain `clicks' at times $t_\text{L}, t_\text{R}$ separately; \RN{4}: apply a second local $\frac{\pi}{2}$-pulses on LHS and RHS atoms respectively \cite{SuppInfo}.
\par
\begin{figure}[t!]
 \centering
\begin{tabular}{l}
\includegraphics[trim = 0mm 0mm 0mm 0mm, clip, width=8.5cm]{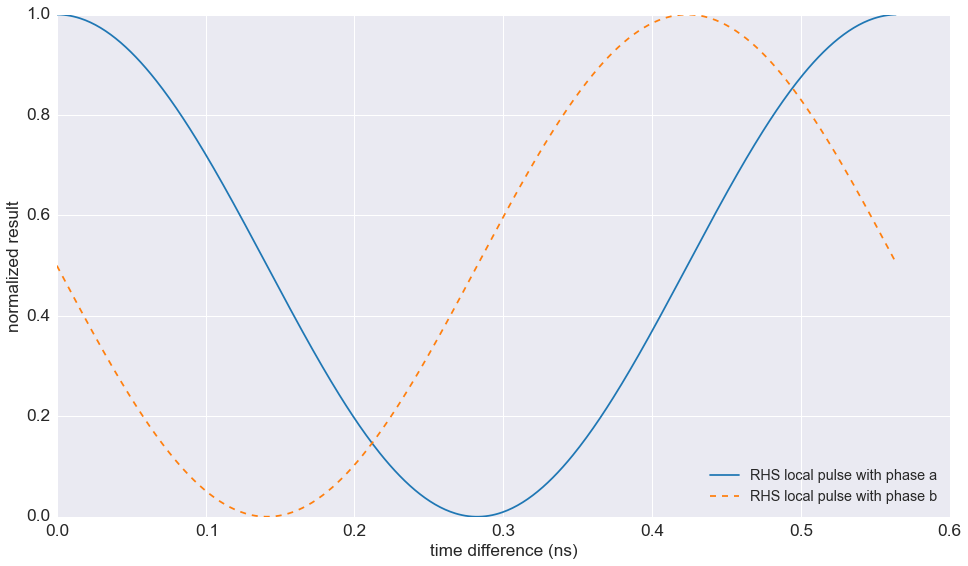}
\end{tabular}
\linespread{1} 
\caption{Numerical simulation for an Bell-type interference experiment of Eq.\eqref{2atom_2spp_heralded1b}, where one more local $\frac{\pi}{2}$-pulse for $|g1\rangle \leftrightarrow |g2\rangle$ is applied on both atoms respectively. The plotted curve can be regarded as the probability corresponding to the measurement outcome of projection onto the basis, with respect to the time difference $t_\text{R}-t_\text{L}$. The frequency detuning $\omega_1 - \omega_2$ is chosen to be 1.77 GHz, and the vertical axis is normalized to 1. This is virtually the same as projecting $|\Psi_{2h}\rangle$ of Eq. \eqref{2atom_2spp_heralded1b} onto the separable state $\frac{1}{2}(\langle g1|_\text{L} + i\langle g2|_\text{L}) (\langle g1|_\text{R} + ie^{i\theta_\text{R}}\langle g2|_\text{R})$. For the optional phase $\theta_\text{R}$, phase setting a is 0 while phase setting b is $\pi/4$. Note that if the photon pulse pair of Eq. \eqref{2atom_2spp_initial1a} does not have the frequency-bin entanglement to begin with, no such interference will be observed; in other words, this proposed interferometry can be regarded as measuring the frequency-bin entanglement of the initial photon pulse pair. Also note that no Hong-Ou-Mandel effect of two-photon interference \cite{PhysRevA.85.021803} or post-selection is employed to obtain this interference result.\label{num_siml_interference_fringe}}
\end{figure}

For each atom, the second local $\frac{\pi}{2}$-pulse can be arranged to be in phase with the first one, such that it transforms $1/\sqrt{2}(|g1\rangle + i|g2\rangle)$ into $|g1\rangle$. Then, up to a global phase, the resulted state for the two atoms in the rotating wave frame is
\begin{align}
\label{2atom_2spp_heralded1b}
|\Psi_{2h}\rangle =
\frac{1}{\sqrt{2}}e^{-i\omega_1 t_\text{L}} e^{-i\omega_2 t_\text{R}}
|g1\rangle_\text{L} |g2\rangle_\text{R} \nonumber\\
+ \frac{1}{\sqrt{2}} e^{-i\omega_2 t_\text{L}} e^{-i\omega_1 t_\text{R}}
|g2\rangle_\text{L} |g1\rangle_\text{R},
\end{align}
where the numerical simulation is presented in Fig. \ref{num_siml_interference_fringe}. This process is equivalent to the optical part of a frequency-bin entangled atom-photon system interacting with a remote atom.
\par
Several known sources of decoherence and dephasing will reduce the fidelity of the interaction from Eq. \eqref{1atom_2spp_initial1a} to Eq. \eqref{1atom_2spp_heralded1a}. (1) photon loss during the process; (2) non-adiabatic transition that puts population into state $|e\rangle$; (3) dephasing between $|g1\rangle$ and $|g2\rangle$, which can be due to environment, laser noise, lack of precision in pulse sequence control \cite{SuppInfo, PhysRevLett.97.140501, PhysRevA.89.032516}.
\par
A pure state of two-component frequency-bin correlation or entanglement is a highly idealized special case. When multiple frequency components or mixed state exist in such an atom-photon correlation structure, state tomography as well as the quantum process tomography becomes a daunting task: even a full characterization of a pure state is at least cumbersome \cite{PhysRevA.91.053851}.
\par
In general, if the temporal waveform complexity is ignored, it is reasonable to discuss the atom-photon frequency-bin entanglement represented by a density matrix $\hat{\rho}$ of dimension $(N_\omega N_A)^2$, where $N_\omega$ is the number of involved frequency components and $N_A$ is the number of involved atomic internal states. Those atomic internal states can be the hyperfine ground level states, magnetically shifted Zeeman sublevels, or Rydberg states, etc. The major task is then reduced to construct an entanglement witness which is convenient both conceptually and practically \cite{PhysRevLett.118.110502}.
\par
It is time to make an observation of a pure state which is separable:
\begin{equation}
\label{multi-comp_separable_1a}
|\Psi_{sp}\rangle = 
(\sum_{m=1}^{N_\omega} b_m|\omega_m\rangle_\text{L})
\otimes
(\sum_{n=1}^{N_A} c_n|A_n\rangle_\text{R}),
\end{equation}
where $|\omega_m\rangle$ denotes the photon pulse, $|A_n\rangle$ denotes the atomic internal state and the normalization condition is that $\sum_m |b_m|^2 = 1$ and $\sum_n |c_n|^2 = 1$.
For those $\omega_m$'s whose difference is smaller than the photon detector's frequency response range, if a time-resolved measurement is performed at LHS resulting in a click at time $t_\text{L}$, the rest is a purely linear superposition of atomic internal wavefunctions, up to a global phase, which can be understood as that conditioned on a time resolved measurement of the optical part, the atomic internal wave function does not change.

\par
As a first step, we discuss the witnessing for a specially constrained subspace. That is, a subspace made up by separable pure states with the same single photon wave form $\Tr(\langle t_\text{L}|\hat{\rho}|t_\text{L}\rangle)$, where $\Tr$ is taking trace over all associated atomic internal states $|A_n\rangle$. For simplicity, suppose the two atomic internal states subject to the entanglement witnessing are $|A_g\rangle$ and $|A_r\rangle$ and on RHS the probabilities of arriving at those two states are equal in a local measurement. 
\par
\begin{figure}[t!]
 \centering
\begin{tabular}{l}
\includegraphics[trim = 0mm 0mm 0mm 0mm, clip, width=8.5cm]{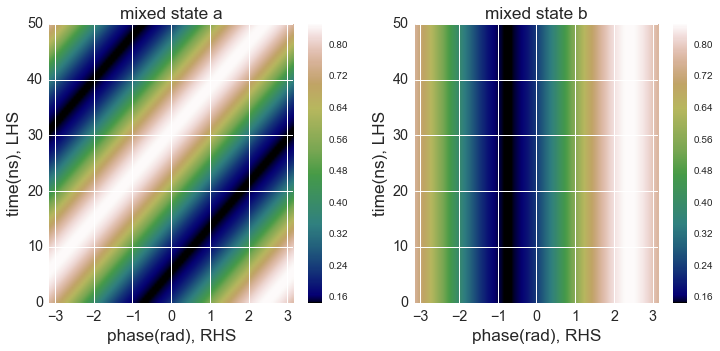}
\end{tabular}
\linespread{1} 
\caption{Numerical simulation of the measurement results for mixed states. All rates are normalized with respect to the same standard. The measurement process is regarded as a projection onto $|t_\text{L}\rangle \otimes \frac{1}{\sqrt{2}}(|g1\rangle_\text{R} + \exp(-i\theta_\text{R})|g2\rangle_\text{R})$, where $t_\text{L}$ is represented by the vertical axis and $\theta_\text{R}$ is represented by the horizontal axis. Mixed state a is chosen to be the equal probability mixture of two pure states $\frac{1}{\sqrt{2}}(|\omega_{1}\rangle_\text{L} |g1\rangle_\text{R} + 
|\omega_{2}\rangle_\text{L}|g2\rangle_\text{R})$ and $\frac{1}{\sqrt{2}}
(|\omega_{1}\rangle_\text{L} |g1\rangle_\text{R} + i|\omega_{2}\rangle_\text{L} |g2\rangle_\text{R})$; while, mixed state b is chosen to be the equal probability mixture of two pure states $|\omega_1\rangle_\text{L} \otimes \frac{1}{\sqrt{2}}(|g1\rangle_\text{R}+|g2\rangle_\text{R})$ and $|\omega_2\rangle_\text{L} \otimes \frac{1}{\sqrt{2}}(|g1\rangle_\text{R}+i|g2\rangle_\text{R})$. The frequency detuning is chosen to be $\omega_1 - \omega_2 = \omega_{g2} - \omega_{g1} =$ 20MHz, and the RHS local projection is assumed to take place always at a fixed time point. \label{num_siml_Entanglement_Witnessing}}
\end{figure}

Assume that the initial time is always fixed. On LHS, a photon detector will perform a time resolved measurement for the photon pulse and receive a click at time $t_L$. On RHS, the local measurement operation is supposed to be equivalent to projecting onto $\frac{1}{\sqrt{2}}(|A_g\rangle + \exp(i\theta) |A_r\rangle)$, always at the same time point. For example, if $|A_g\rangle$ \& $|A_r\rangle$ are already in the rotating wave frame associated with some previous local operations, then the criteria is to keep track of the RHS master clock, similar to the situation of Eq.\eqref{1atom_2spp_heralded1b}. Based upon those settings for the measurement, we can define the following function:
\begin{align}
\label{multi-comp_dm_cond1}
F_c&(t_L, \theta, \hat{\rho}) = 
(\sum_{n=1}^{N_A} \langle A_n | \langle t_\text{L}| \hat{\rho} 
|t_\text{L}\rangle |A_n\rangle)^{-1} \nonumber \\
&\cdot (\langle A_g| + \exp(-i\theta) \langle A_r|) \langle t_\text{L}|
\hat{\rho} |t_\text{L}\rangle (|A_g\rangle + \exp(i\theta) |A_r\rangle).
\end{align}
Henceforth the criteria of entanglement witness is then as straightforward as whether $F_c (t_L, \theta)$ is a function of $\theta$ only, see Fig. \ref{num_siml_Entanglement_Witnessing} for a numerical example. A succinct proof can be sketched here. Suppose that $\hat{\rho}_{sp1}$ and $\hat{\rho}_{sp2}$ are the density matrices for two different separable states,. According to Eq.\eqref{multi-comp_separable_1a}, both $F_{c}(t_\text{L}, \theta, \hat{\rho}_{sp1})$ and $F_{c}(t_\text{L}, \theta, \hat{\rho}_{sp1})$ are independent of $t_\text{L}$. Moving on the to the convex combination $\beta_1 \hat{\rho}_{sp1} + \beta_2 \hat{\rho}_{sp2}$ where $\beta_1 + \beta_2 = 1, \beta_1, \beta_2 >0$; now we have $F_{c}(t_\text{L}, \theta, \beta_1 \hat{\rho}_{sp1} + \beta_2 \hat{\rho}_{sp2}) = \beta_1 F_{c1}(t_\text{L}, \theta) + \beta_2 F_{c2}(t_\text{L}, \theta)$, and is again independent of $t_\text{L}$.
\par
Even though the ability of entanglement witness based upon Eq. \eqref{multi-comp_dm_cond1} is severely limited because it applies to a subspace spanned by very special separable states, it contains some advantages. At least, it is most effective to judge the entanglement between `frequency component correlation', namely mixed state made up by pure states $|\omega_j\rangle |A_j\rangle$, and the multi-component frequency-bin entanglement, namely $\sum c_j |\omega_j\rangle |A_j\rangle$.
\par
The next step is to push forward towards a witnessing device with the same philosophy, that is applicable to a convex set of many more separable states. For simplicity, assume that on RHS the two atomic internal electronic states of interest are $|A1\rangle$ and $|A2\rangle$. And then we need a few definitions for global and local measurement outcomes:
\begin{subequations}
\label{dm_FEW_defs}
\begin{align}
\label{dm_FEW_def_1a}
K_\text{LR}(\hat{\rho}, t, \theta) 
&=\frac{1}{\sqrt{2}}(\langle A1|_\text{R} + e^{-i\theta_\text{R}} \langle A2|_\text{R})\langle t |_\text{L}
\cdot \hat{\rho}  \nonumber\\
&\, \cdot |t\rangle_\text{L}
\frac{1}{\sqrt{2}}(|A1\rangle_\text{R} + e^{i\theta_\text{R}} |A2\rangle_\text{R});\\
\label{gate_dm_sp_1b}
K_\text{L} (\hat{\rho}, t_s)
&= \Tr\{ \langle t_s |_\text{L} \cdot \hat{\rho} \cdot | t_s \rangle_\text{L}\};\\
K_\text{R} (\hat{\rho}, \theta_\text{R}) 
&=\int \frac{1}{\sqrt{2}}(\langle A1|_\text{R} + e^{-i\theta_\text{R}} \langle A2|_\text{R})\langle t |_\text{L}
\cdot \hat{\rho}  \nonumber\\
&\, \cdot |t\rangle_\text{L}
\frac{1}{\sqrt{2}}(|A1\rangle_\text{R} + e^{i\theta_\text{R}} |A2\rangle_\text{R})\,dt;
\end{align}
\end{subequations}
where the definition of those projections are the same as before. Moreover, we can define the associated Fourier transforms for Eq. \eqref{dm_FEW_defs}:
\begin{subequations}
\label{dm_FEW_def_1b}
\begin{align}
\mathcal{F}[K_\text{LR}](\hat{\rho}, \omega_s, n_\text{R}) &=
\frac{1}{2\pi T} \iint K_\text{LR}(\hat{\rho}, t_c, \theta) \nonumber\\
& \qquad \cdot e^{-in_\text{R}\theta_\text{R}} e^{-i\omega_s t_s} d\theta_\text{R} dt_s;
\\
\mathcal{F}[K_\text{L}](\hat{\rho}, \omega_s) &=
\frac{1}{T}\int_T K_\text{L} (\hat{\rho}, t_s) e^{-i\omega_s t_s} dt_s;
\\
\mathcal{F}[K_\text{R}](\hat{\rho}, n_\text{R}) &=
\frac{1}{2\pi} \int_{2\pi} K_\text{R} (\hat{\rho}, \theta_\text{R}) e^{-in_\text{R}\theta_\text{R}} d\theta_\text{R};
\end{align}
\end{subequations}
where in practice $\omega_s$ takes discrete values $2\pi m/T, m = 0, 1, 2 \ldots$, where $T$ is the total time of the measurement window. 
\par
For a separable pure state $\hat{\rho}_{sp}$, a special condition holds as the following:
\begin{equation}
\label{dm_sp_2a}
K_\text{LR}(\hat{\rho}_{sp}, t_s, \theta) =
K_\text{L} (\hat{\rho}_{sp}, t_s) \cdot K_\text{R} (\hat{\rho}_{sp}, \theta_\text{R}),
\end{equation}
such that the Fourier transform obey:
\begin{align}
\label{dm_sp_2_Fourier}
&\iint K_\text{LR}(\hat{\rho}_{sp}, t_c, \theta) 
e^{-in_\text{R}\theta_\text{R}} e^{-i\omega_s t_s} d\theta_\text{R} dt_s
= \nonumber\\
&\, \int K_\text{L} (\hat{\rho}_{sp}, t_s) e^{-i\omega_s t_s} dt_s
\int_{2\pi} K_\text{R} (\hat{\rho}_{sp}, \theta_\text{R}) e^{-in_\text{R}\theta_\text{R}} d\theta_\text{R}.
\end{align}
\par
Fixing the range of interest for $\omega_s, n_\text{R}$. Based upon the above observations, we construct an entanglement witness for the set of separable states formed by the collection of all convex combinations of pure product states satisfy $|\mathcal{F}[K_\text{L}](\hat{\rho}_{sp}, \omega_s)| \leq \epsilon_\text{L}$ or $|\mathcal{F}[K_\text{R}](\hat{\rho}_{sp}, n_\text{R})| \leq \epsilon_\text{R}$ for some parameters $0 \leq \epsilon_\text{L}, \epsilon_\text{R} \leq 1$. Then, for a general state $\hat{\rho}$, given following set of two inequalities:
\begin{subequations}
\label{dm_sp_EW}
\begin{align}
|\mathcal{F}[K_\text{LR}](\hat{\rho}, \omega_s, n_\text{R})|
\leq \epsilon_\text{L};
\\
|\mathcal{F}[K_\text{LR}](\hat{\rho}, \omega_s, n_\text{R})|
\leq \epsilon_\text{R};
\end{align}
\end{subequations}
where if one of the inequalities are violated, then the mixed state or pure state represented by $\hat{\rho}$ is not separable. The proof is straightforward; namely, knowing that $\hat{\rho}_{sp1}$ and $\hat{\rho}_{sp2}$ are pure states, then the inequalities of Eq.\eqref{dm_sp_EW} for $\beta_1 \hat{\rho}_{sp1} + \beta_2 \hat{\rho}_{sp2}$ with $\beta_1 + \beta_2 = 1, \beta_1, \beta_2 >0$ cannot be violated \cite{SuppInfo}.
\par
In conclusion, we have proposed a particular form of frequency-bin type entanglement between photon pulses and atom's internal energy levels. We have shown that it comes naturally from the interaction between frequency-bin entangled photon pulses and single atoms. We have also studied its fundamental entanglement properties and constructed straightforward entanglement witnessing tools.

\begin{acknowledgments}
The authors acknowledge support from NSFC and NUDT. 
The authors gratefully acknowledge Professor Mark Saffman for his help which makes this work possible. The authors thank Dr. Shuyu Zhou for carefully reviewing the manuscript and enlightening discussions.
\end{acknowledgments}

\bibliographystyle{apsrev4-1}

\renewcommand{\baselinestretch}{1}
\normalsize

\bibliography{ysref}
\end{document}